\documentclass[a4paper,12pt]{article}
\usepackage[dvips]{graphicx}

\usepackage{epsfig}

\usepackage{amsmath}

\newcommand{\D}{\ensuremath{\mathrm{d}}}

\begin{document}

\begin{center}
  \quad\\
{\Large Symmetry breaking through a sequence
of transitions \\[1ex] in a driven diffusive system}
 \vskip1cm {\large
M. Clincy$^1$, M. R. Evans$^1$, D. Mukamel$^2$}
\end{center}
\vskip1cm
\begin{center}
$^1$ Department of Physics and Astronomy,\\ University of Edinburgh,
       Mayfield Road, Edinburgh EH9 3JZ, U.K. \\[2ex]
$^2$  Department of Physics of Complex Systems, \\The Weizman Institute of
       Science,
 Rehovot 76100, Israel
\end{center}
\begin{center}
\today
\end{center}
\vskip1cm
\begin{abstract}
  In this work we study a two species driven diffusive system with
  open boundaries that exhibits spontaneous symmetry breaking in one
  dimension. In a symmetry broken state the currents of the two
  species are not equal, although the dynamics is symmetric. A mean
  field theory predicts a sequence of two transitions from a strongly
  symmetry broken state through an intermediate symmetry broken state
  to a symmetric state. However, a recent numerical study has
  questioned the existence of the intermediate state and instead
  suggested a single discontinuous transition.  In this work we
  present an extensive numerical study that supports the existence of
  the intermediate phase but shows that this phase and the transition
  to the symmetric phase are qualitatively different from the
  mean-field predictions.
\end{abstract}
\vfill

\section{Introduction}
Nonequilibrium systems may be defined as systems evolving according to
dynamics that are not constructed from any detailed balance condition.
Thus their steady states are not described by Gibbs-Boltzmann
statistical mechanics. In particular, such a steady state is
characterised by nonvanishing currents of probability between
configurations.  These
nonequilibrium steady states have been the subject of much attention
for some years now and their properties are of fundamental interest
\cite{David}. For example, generic long-range correlations and
nonequilibrium phase transitions, even in one dimension, may be
exhibited \cite{SZ,MD,Evans}; such behaviour is in contrast to that of
equilibrium steady states.

One class of nonequilibrium systems are driven diffusive systems (DDS).
As well as probability currents existing these systems also have a
mass current of a conserved quantity driven through the system.  DDS
lend themselves to the modelling of a variety of phenomena such as
super-ionic conductors, traffic flow and  biophysical transport
problems \cite{David,SZ}.

Perhaps the simplest DDS is the asymmetric simple exclusion process
(ASEP). Here particles hop in a preferred direction on a
one-dimensional lattice with hard-core exclusion (at most one particle
can be at any given site).  The open system where particles are
injected and extracted at the boundaries was first studied by
Krug\cite{Krug} and boundary-induced phase transitions shown to be
possible.  Since then our understanding of phase transitions in
one-dimensional DDS has deepened through exact solutions of this model
being worked out \cite{DDM}--\cite{Karimipour} and further examples
being studied.  Of particular interest have been phase
transitions manifesting spontaneous symmetry breaking
\cite{EFGM,GLEMSS,AHR,AH,PP}, phase transitions and phase separation induced
by defect particles or defect sites
\cite{Evans96,KF,Mallick,DE99,SasamotoPRE,TB}, 
phase separation in multicomponent systems \cite{EKKM,AHRJPA}
and further examples of boundary-induced phase transitions
\cite{PS,AS}.  Clearly the presence of conserved quantities and a drive,
leading to non-zero current $j$ is crucial to all these phase
transitions.

Exact results have provided us with a particularly good understanding
of the ASEP with open boundary conditions.  Let us summarise some of
the results here.  The open system is defined on a lattice of $N$
sites where at the left boundary a particle is introduced with rate
$\alpha$ if that site is empty, and at the right boundary site any
particle present is removed with rate $\beta$.  These boundary
conditions force a steady state current of particles $j_N$ through the
system. Phase transitions occur when the current in the thermodynamic
limit, $j= \lim_{N{\to}\infty} j_N$, exhibits non-analyticities.  The
steady state of this system was solved exactly first for the totally
asymmetric hopping case \cite{DEHP,SD}.  The phase diagram comprises
three phases: a high-density phase where the system is controlled by
the low exit rate $\beta$ ($\beta < 1/2; \alpha>\beta)$, the current
takes the value $j=\beta(1-\beta)$ and the bulk density is $\rho =
(1-\beta)$; a low-density phase where the system is controlled by a
low injection rate $\alpha$ ($\alpha < 1/2 ; \alpha < \beta$), the
current is $j=\alpha(1-\alpha)$ and the bulk density is $\rho=\alpha$;
a maximal-current phase where both $\alpha,\beta > 1/2$ and the
current is $j=1/4$ and the bulk density $\rho=1/2$.  The exact
solution has been generalised to partially asymmetric hopping and
retains the same three generic phases \cite{SasamotoJPA,BECE}.  Indeed,
the qualitative phase diagram appears robust for one-dimensional,
open, driven systems \cite{KSKS,HKPS}.

The phase transitions from the high density or low density phases to
the maximal-current phase are both {\em continuous}---the density is
continuous at the transition but has a discontinuity in its first
derivative and the current has a discontinuity in its second
derivative.  However the transition from the high density to the low
density phase is {\em discontinuous}---the density is discontinuous
and the current has a discontinuity in its first derivative at the
transition.  On the coexistence line between the high and low density
phases the system consists of a low density region adjacent to the
left boundary and a high density region adjacent to the right
boundary.  The currents within the two domains are equal (otherwise
one phase would be driven out of the system) and domains are separated
by a shock, i.e. a sharp domain wall that performs diffusive motion due
to fluctuations in the number of particles entering and leaving the
system \cite{DEMall}.

The above observations suggest an analogy between the role of the
current for a driven system and the role of the chemical potential in
an equilibrium system.  That is, whereas in equilibrium the condition
for phase coexistence is that the chemical potential of the two phases
must be equal, in the present case one must have that the currents in
the two phases are equal.  Indeed, the exact solution \cite{DEHP}
revealed that the current is expressed as the ratio of partition sums
for systems of size $N{-}1$ and $N$ yielding an explicit
identification of $j$ as a fugacity (i.e.  the exponential of a
chemical potential).

A further kind of boundary-induced phase transition,
manifesting spontaneous symmetry breaking,
is found when the model is generalised to two oppositely
moving species of particle: one species is injected at the left, moves
rightwards and exits at the right; the other species is injected at
the right, moves leftwards and exits at the left \cite{EFGM}.
Intuitively one can picture the system as a narrow road bridge: cars
moving in opposite directions can pass each other but cannot occupy the same
space.  The model has a left-right symmetry when the injection rates
and exit rates for the two species of particles are symmetric.
However for low
exit rates ($\beta$) this symmetry is broken
and   the lattice is dominated by one of the species at
any given time. This implies that the short time averages of
currents and  bulk densities of the
two species of particles are no longer equal.
Over longer times the system flips between
the two symmetry-related states.
In the $\beta \to 0$ limit the mean flip time between
the two states  has been calculated analytically and
shown to diverge exponentially with system size \cite{GLEMSS}.  Thus
the `bridge' model provided a first example of spontaneous symmetry
breaking in a one dimensional system.

Of particular interest is the transition from the low $\beta$ regime
where the symmetry is broken to the high $\beta$ regime where the
steady state is symmetric.  A mean-field theory \cite{EFGM}(to
be reviewed in section~\ref{mft}) predicts the following
scenario in terms of the phases of the single species model.  For low
$\beta$ one species of particles (the dominant one) is in a high
density phase whereas the other is in a low density phase. This will
be referred to as the {\it hd/ld} phase. As $\beta$ is increased a
discontinuous transition occurs to another asymmetric phase where both
particles are in single species low density phases ({\it ld/ld} phase)
i.e. the density of each species is less than 1/2 but the densities
are different. As $\beta$ is increased further a continuous transition
occurs to a symmetric phase wherein the species have the same density
with value less than 1/2 ({\it ld} phase).

Thus, in this scenario, there is a sequence of {\it two} transitions
with the {\it ld/ld} phase appearing as the intermediate phase.  The
intermediate phase occupies a very narrow region of the mean field
phase diagram, nevertheless in \cite{EFGM}, Monte Carlo simulations
provided some evidence supporting its existence.

However, Arndt {\it et al} \cite{AHR} have argued on the basis of numerical
studies that there is a {\it single} discontinuous transition and
that the {\it ld/ld} phase does not exist. 
In order to draw their conclusions Arndt {\it et al} used the concept
of a nonequilibrium free energy.  This simply involves calculating the
probability distribution of the particle densities in the steady
state.  Arndt {\it et al} then converted this into a `free energy density' as
a function of particle densities by taking the logarithm of the
distribution and dividing by $N$.

In order to clarify the discrepancy between the results of \cite{EFGM}
and \cite{AHR} we further investigate in this work the transition from
the asymmetric state to the symmetric state.
We present extensive numerical studies and theoretical arguments
that suggest that the intermediate {\it ld/ld} phase does indeed exist.
However, our numerics indicate that the sequence of transitions
is distinct from both the mean-field prediction and the scenario
proposed in \cite{AHR}. Instead, we observe a third scenario in which
as $\beta$ is increased a
discontinuous transition occurs 
from the {\it hd/ld} to the {\it ld/ld} phase, then a further
discontinuous transition  occurs from the {\it ld/ld}
to the symmetric phase.

The two first order transitions displayed by this model are found to 
be of a different nature. In the transition between the {\it hd/ld} and the
{\it ld/ld} phases, the two phases have the same currents, and thus
they may {\it coexist} in the same system with one fraction of the
system occupied by one phase and the rest of the system by the other.
This is closely analogous to ordinary first order transitions in equilibrium
where the phases involved in the transition have the same chemical
potential. On the other hand the first order transition from the asymmetric 
{\it ld/ld} to the symmetric phase is found to be of a different nature, which
has no equilibrium analogue. Here the currents of the two phases are not
equal to each other and thus the two phases cannot coexist in the same
system for a long time.

The layout of the paper is as follows.
In Section~2 we define the model and review the mean-field theory.
In Section~3 our numerical simulations are presented,
in particular we provide numerical evidence for two discontinuous
transitions.
In Section~4 we discuss a simplistic `blockage' picture that
captures some features of the simulations.
In Section~5 we conclude and resolve the discrepancies
between our conclusions and those of \cite{AHR}.

\section{Model definition and mean-field theory}
We now define the bridge model
described in the introduction.
We indicate the presence of
a hole by `0', and of a positive or negative particle by `$+$' 
and `$-$' respectively. The site 
is denoted by an index. The possible configuration changes that may 
occur in time $\D t$ can be described 
as follows:

\noindent 
left end:
\begin{equation}
\begin{array}{lcl}
(0)_1 \to (+)_1 & \mbox{with probability} & \alpha\; \mbox{d}t\\[1ex]
(-)_1 \to (0)_1 & \mbox{with probability} & \beta \; \mbox{d}t
\end{array}\label{left}
\end{equation}
right end:
\begin{equation}
\begin{array}{lcl}
(0)_N \to (-)_N & \mbox{with probability} & \alpha \; \mbox{d}t\\[1ex]
(+)_N \to (0)_N & \mbox{with probability} & \beta\;  \mbox{d}t
\end{array}\label{right}
\end{equation}
bulk ($1<i<N$):
\begin{equation}
\begin{array}{lcl}
(+)_i(0)_{i+1} \to (0)_i(+)_{i+1} & \mbox{with probability} & \mbox{d}t\\[1ex]
(0)_i(-)_{i+1} \to (-)_i(0)_{i+1} & \mbox{with probability} & \mbox{d}t\\[1ex]
(+)_i(-)_{i+1} \to (-)_i(+)_{i+1} & \mbox{with probability} & \mbox{d}t
\end{array}\label{bulk}
\end{equation}

Thus the two types of particles---referred to as `positive' and
`negative' particles although they exhibit only hard-core
interaction---hop with rate 1 in opposite directions on a lattice of
$N$ sites. Positive particles are injected at the left-hand side with
rate $\alpha$, if the first site is empty, and removed from the
right-hand boundary with rate $\beta$ if the last site is occupied.
With respect to the two types of particles the model is symmetric,
thus negative particles enter the system on the right-hand side with
rate $\alpha$, and leave on the left-hand side with rate $\beta$.

In the following we will consider only the case $\alpha=1$.

\subsection{Mean-field theory}\label{mft}
Here we review the mean-field theory and associated phases
\cite{EFGM}. The mean-field theory is implemented
by approximating two-point correlation functions by 
products of one-point correlations functions.

Let $p_i$ ($m_i$) denote the average positive (negative) particle density 
at site $i\in \{1,2,...,N\}$.
In a stationary state, 
the positive (negative) particle currents 
$j^{+}$ ($j^{-}$) 
through the system are site independent
and satisfy, in the mean-field theory,
\begin{eqnarray}
j^+ &=& p_i\left(1 - p_{i+1}\right)\label{jp_bulk} \\
j^- &=& m_{i+1}\left(1 - m_i\right)\label{jm_bulk}
\end{eqnarray}
for the bulk ($1<i<N$) and
\begin{eqnarray}
j^+ &=& \left(1 - p_1 - m_1\right) =\beta p_N \label{jp_boundary}\\
j^- &=& \beta m_1=  \left(1 - p_N - m_N\right)\label{jm_boundary}
\end{eqnarray}
for the boundaries.
We will denote the limiting bulk densities far away from the boundaries
as $p,m$ so that in the limit of large $N$
\begin{eqnarray}
j^+ &\to& p\left(1 - p\right)\label{jplus_bulk} \\
j^- &\to& m\left(1 - m\right)\label{jminus_bulk}\;.
\end{eqnarray}

Notice that the two systems of particles are only coupled 
by the boundary conditions (\ref{jp_boundary},\ref{jm_boundary}).
The mean-field theory solves  for the currents through
the definition of effective boundary rates
\begin{eqnarray}
  \alpha^+ & = & \frac{1 - p_1 - m_1}{1 - p_1}
 = \frac{j^+}{j^+ + j^-/\beta} \label{aplusdef}\\
  \alpha^- & = & \frac{1 - p_N - m_N}{1 - m_N}
 = \frac{j^-}{j^- + j^+/\beta} \label{aminusdef}\\
  \beta^+ & = & \beta^- = \beta
\end{eqnarray}
Then the equations for  the currents of each species can  be solved
self-consistently by using the
solutions of the mean-field theory
for the single  species model \cite{DDM}.

Depending on the feeding and output rates $\alpha^S$ and $\beta^S$ 
(superscript ``S'' denoting the single species quantities) there exist three different phases in the single species model:
\begin{itemize}
\item The maximal-current or power-law phase with current $j^S = 1/4$ for $\alpha^S\ge 1/2$ and $\beta^S\ge 1/2$. The bulk density is $1/2$, the density profile obeys a power law.
\item The low-density phase for $\alpha^S < \beta^S$ and $\alpha^S < 1/2$. The current is $j^S = \alpha^S(1- \alpha^S)$ and the bulk density is equal to $\alpha^S$.
\item The high-density phase for $ \beta^S <\alpha^S$ and $\beta^S< 1/2$. The current is $j^S = \beta^S(1- \beta^S)$ and the bulk density equals $1-\beta^S$.
\end{itemize}

Considerations concerning which of these phases are compatible with
each other lead to the identification of four
phases for the two species system\cite{EFGM}. 
\begin{itemize}
\item Symmetric phases:
in these phases $j^+ = j^-$, hence from (\ref{aplusdef},\ref{aminusdef})
\begin{equation}
\alpha^S = \alpha^+ = \alpha^- = \frac{\beta}{1 + \beta}\label{sym}
\end{equation}
        \begin{itemize}
        \item Maximal  Current symmetric phase:
both species are in the maximal single species phase. Thus
 $j^+ =j^- = 1/4$ and $p=m=1/2$. The phase exists  for $\alpha^S \ge
 1/2$ which corresponds to $\beta \ge 1$. 

        \item Low-density symmetric phase:
both species are in the low density single species phase.
Thus $j^+=j^-= \beta/(1+\beta)^2$ and $p=m=\beta/(1+\beta)$.
The phase exists for $\beta \le 1$. 
        \end{itemize}
\item Asymmetric phases: in these phases $j^+ \neq j^-$ 
(it will be assumed that the positive particles are the majority
species
i.e. $j^+ > j^-$):
        \begin{itemize}
        \item High-density/low-density phase:
positive particles
are in a high density  phase and negative particles in a low density
phase thus
        \begin{eqnarray}
        j^+ &=& \beta \left(1 - \beta\right)\label{hlmaj_curr}\\
        j^- &=& \alpha^- \left(1- \alpha^-\right)\label{hlmin_curr}
        \end{eqnarray}
where
        \begin{equation}
        \alpha^- = 1 - \sqrt{1-\beta}\;.
        \end{equation}
        The bulk densities are 
\begin{equation}
p= 1-\beta\quad\mbox{and}\quad m= 1 -
        \sqrt{1-\beta}\;. 
\label{mfdensity}
\end{equation}
         This phase is found for
        $\alpha^{+} > \beta$ which yields using (\ref{aplusdef},
\ref{hlmaj_curr},\ref{hlmin_curr}) $\beta < 0.329$
        \item Low-density/low-density asymmetric phase:
positive particles
and negative particles are in distinct low density
phases thus
        \begin{eqnarray}
        j^+ &=& \alpha^+ \left(1- \alpha^+\right)\\
        j^- &=& \alpha^- \left(1- \alpha^-\right)
        \end{eqnarray}
        where $\alpha^+ \neq\alpha^-$. 
Some calculation yields for the densities 
 \begin{eqnarray}
       p= \alpha^+ &=& \frac{1}{2(1-\beta)}\left[
  (1-2\beta) + \left( \frac{1-3\beta}{1+\beta}\right)^{1/2}\right]
\label{pmft}\\
        m=\alpha^- &=& \frac{1}{2(1-\beta)}\left[
  (1-2\beta) - \left( \frac{1-3\beta}{1+\beta}\right)^{1/2}\right]
\label{mmft}
        \end{eqnarray}
The condition for this phase is $\alpha^{+} < \beta$
i.e. $\beta < 0.329$. There is a continuous transition
to the {\it ld} symmetric phase at $\beta=1/3$
\end{itemize}
\end{itemize}
Thus, as we increase $\beta$ from zero the mean-field theory predicts
the following sequence of transitions:
at  $\beta =0.329$ a discontinuous transition from {\it hd/ld} to
{\em ld/ld};
at $\beta = 1/3$ a continuous transition from the
{\it ld/ld} asymmetric to the low-density symmetric phase;
at $\beta =1$ a continuous transition from the
{\it ld}  phase to the maximal current  phase.

\section{Simulations}

We have carried out various numerical simulations to investigate the 
transition from the {\it hd/ld} to the {\it ld} symmetric phase to 
gain some insight into whether a second asymmetric phase is present. 
A standard random sequential updating procedure  is used to simulate the
dynamics Eqns. (\ref{left}--\ref{bulk}).

First of all, we followed the idea of Arndt {\it et al} \cite{AHR} 
and simulated the probability distribution $P(p,m)$ of the steady state as a 
function of  both positive and negative particle density
$p$ and $m$ and for various $\beta$.
In a simulation  the  positive and
negative particle densities $p$ and $m$ are evaluated every
ten Monte Carlo steps to built up a histogram for the probability 
distribution $P(p,m)$.

\begin{figure}[htbp]
        \begin{center}
              
                \includegraphics[height=96mm]{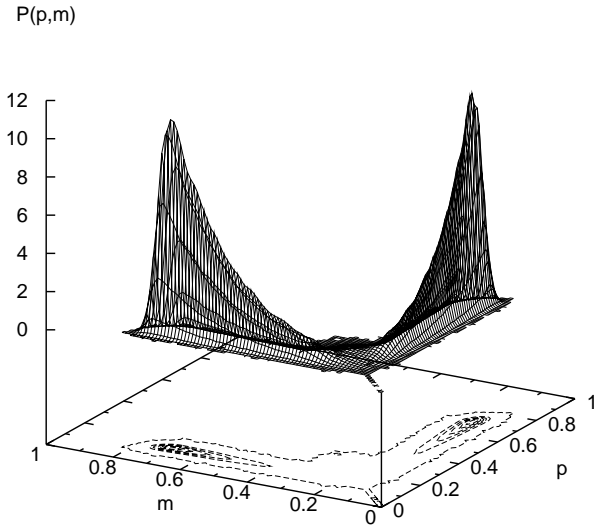}
\end{center}
\caption{A 3-d plot of the probability density $P(p,m)$ as a function
of the  particle densities $m,p$ for $N=500$, $\beta=0.265$.
A contour plot is shown  projected onto the $m-p$ plane.}
\label{beta_0265}
\end{figure}

\begin{figure}[htbp]     
        \begin{center}
                 \includegraphics[height=96mm]{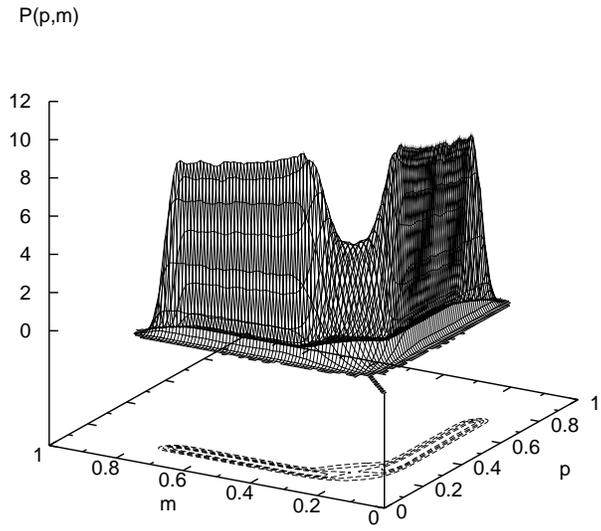}
\end{center}
\caption{$P(p,m)$ as in Fig.~\ref{beta_0265} for $N=500$, $\beta=0.274$, different scale than Figures \ref{beta_0265}, \ref{beta_0280}--\ref{beta_0290}}
\label{transition}
\end{figure}

\begin{figure}[htbp]     
        \begin{center}
               
                \includegraphics[height=96mm]{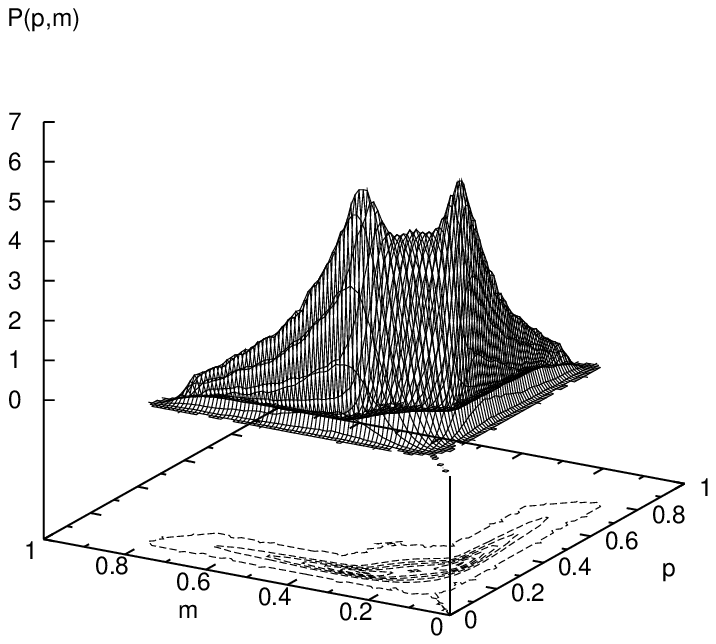}
\end{center}
\caption{$P(p,m)$ as in Fig.~\ref{beta_0265} for $N=500$, $\beta=0.280$}
\label{beta_0280}
\end{figure}

\begin{figure}[htbp]     
        \begin{center}
               
                \includegraphics[height=96mm,angle=-90]{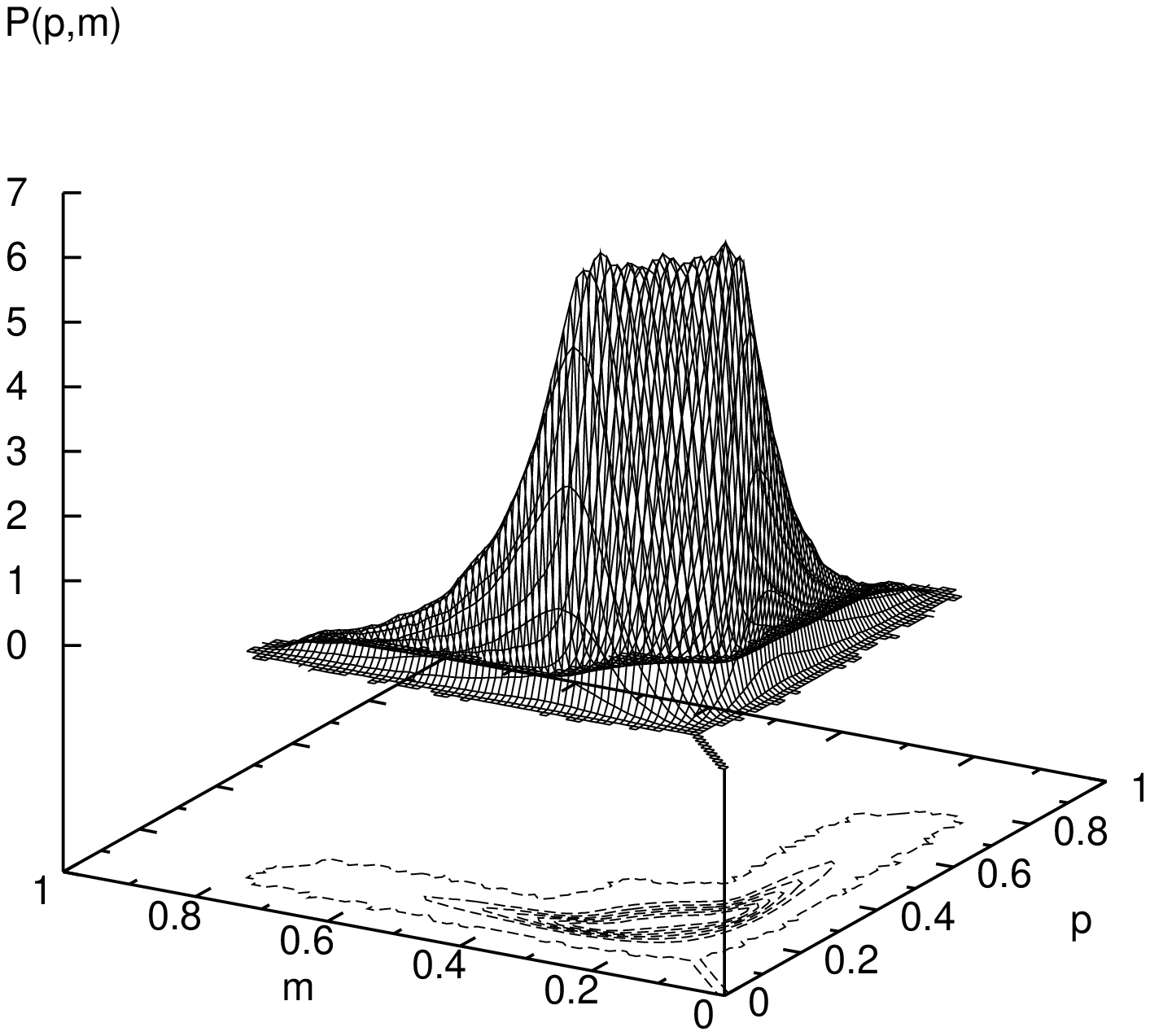}
\end{center}
\caption{$P(p,m)$ as in Fig.~\ref{beta_0265} for $N=500$, $\beta=0.284$}
\label{beta_0284}
\end{figure}

\begin{figure}[htbp]     
        \begin{center}
               
                \includegraphics[height=96mm]{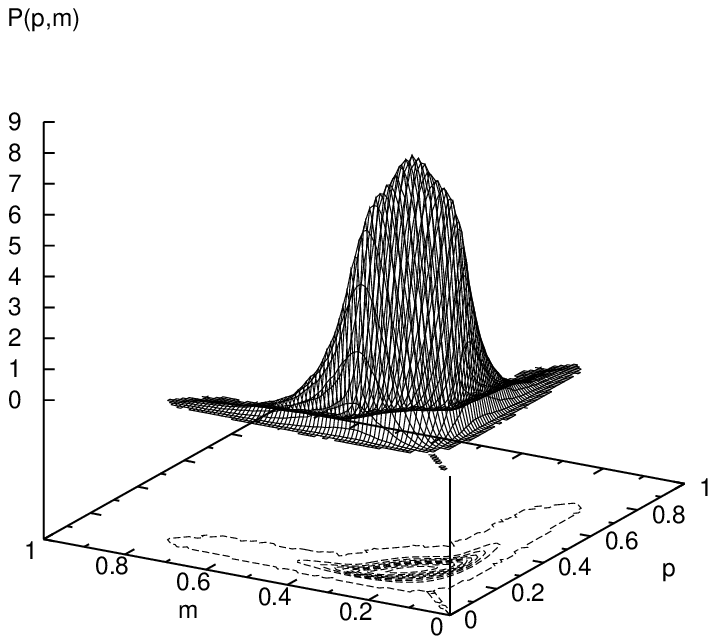}
\end{center}
\caption{$P(p,m)$ as in Fig.~\ref{beta_0265} for $N=500$, $\beta=0.290$}
\label{beta_0290}
\end{figure}

As shown in Figures \ref{beta_0265} to \ref{beta_0290} where $P(p,m)$
is plotted for a system size of $N=500$ sites and $\beta = 0.265,
0.274, 0.280, 0.284$ and $0.290$ the three different phases
qualitatively predicted by mean-field theory can be
found. Quantitatively, the values differ from the mean-field results
which predict the transition from the {\it hd/ld} to the {\it ld/ld}
for $\beta=0.329$ and the transition to the symmetric phase for
$\beta=1/3$.


In Fig. 1 $\beta = 0.265$ and the system is in the {\it hd/ld} phase.
The probability distribution shows two well-separated peaks at $(p,m)
\approx (0.74,0.14)$ and $(p,m) \approx (0.14,0.74)$ which correspond
to the densities $\rho_{\rm majority} = 1-\beta$ and $\rho_{\rm
  minority} = 1-\sqrt{1-\beta}$ predicted by mean-field theory in
section \ref{mft}. A transition occurs at $\beta\approx 0.274$
(Fig.~\ref{transition}). One observes the `boomerang' shape described
by Arndt {\it et al }: in the two arms of the boomerang of the
probability distribution the density of one particle type is constant
and the density of the other fluctuates strongly.  This behaviour
corresponds to coexistence of the
{\it hd/ld} and {\it ld/ld} phases in the same system with a wandering domain
wall or shock that is equally likely to be at any position
separating the regions of different
density.  However, there is
a clear saddle point separating the two arms of the boomerang.  This is
in contrast to the first order transition proposed by Arndt {\it et
  al.}  in which $P(p,m)$ is supposed to be constant along the whole
structure.  At $\beta = 0.280$ (Fig.~3), we observe
evidence for a {\it ld/ld} asymmetric phase. The probability
distribution shows two peaks at values which are $<1/2$ for both $p$
and $m$ confirming the existence of a {\it ld/ld} asymmetric phase.
However note that the saddle point separating the two peaks itself has 
a high probability.  
When $\beta$ is increased to
0.284 (Fig.~4) the probability appears to have a flat  top
indicating that one is at the transition 
to the symmetric phase. Finally,
For $\beta = 0.29$  the system is in the
symmetric phase with $p=m = 0.22 \simeq \beta/(1+\beta)$.

The maxima in the probability distribution, corresponding to a second
symmetry broken phase, are clearly visible for $\beta = 0.280$, but
there remain some questions about the nature of this {\it ld/ld}
phase. First of all, the location of the peaks is not correctly
described by mean-field theory; the minority density is $\rho_{\rm
  minority} \simeq 1-\sqrt{1-\beta}$ as for the {\it hd/ld} phase, but
the majority species is found to be present with density $\rho_{\rm
  majority} \simeq \beta$ which does not agree
with the mean-field value. Furthermore, it is striking that the maxima
are not only separated by a very high saddle, but also have long tails 
in the {\it hd/ld} region of the distribution that are a vestige of the
transition from the {\it hd/ld} phase (Fig.~2).

To gain a deeper understanding of these features, we explicitly
illustrate the time evolution of the densities at $\beta=0.275$ close
to the transition from the {\it hd/ld} phase (Figs. 6--8).  The
sequence shows different periods of a single simulation run.  The
positive and negative particle densities $p$ and $m$ and their
difference are plotted.  Different types of behaviour are clearly
observed as time evolves.  In Fig.~6 the simulation begins with
densities corresponding to the arm of the boomerang: the density of
the majority phase fluctuates strongly, while that of the minority
phase is relatively constant.  In Fig.~7 the simulation moves into a
short period of symmetric densities corresponding to the {\it ld}
phase then progresses to a symmetry broken period corresponding to the
{\it ld/ld} phase.  Finally Figure~\ref{oscil} illustrates curious
oscillations in the densities about the symmetric value that can
sometimes be observed.

Thus in the {\it ld/ld} region there appear to be different 
behaviours competing
 for the true phase behaviour. One might expect this
to be a finite size effect and that for $N\to \infty$ one
behaviour  dominates. We will provide evidence for this below.

Another consequence of competing behaviours is that they do not
coexist in a system, rather the system switches from one behaviour to
another.  This implies that although the transition to from the {\it
  ld/ld} phase to the symmetric is discontinous there is no
coexistence between the phases. This is in contrast to the
discontinous transition from the {\it hd/ld} to the {\it ld/ld} phase
where there is coexistence.
One can also rule out coexistence at the {\it ld/ld} to symmetric
phase transition on theoretical grounds: it  is not possible 
for a phase with symmetric currents to coexist with a phase with
asymmetric currents in the same system.
Curiously, although coexistence does not occur, 
figure~\ref{beta_0284}, appears to have a
flat-topped 
distribution.
It would be interesting to see whether this is just a finite  size effect.

\begin{figure}[htbp]
        \begin{center}
                \includegraphics[height=78mm,angle=-90]{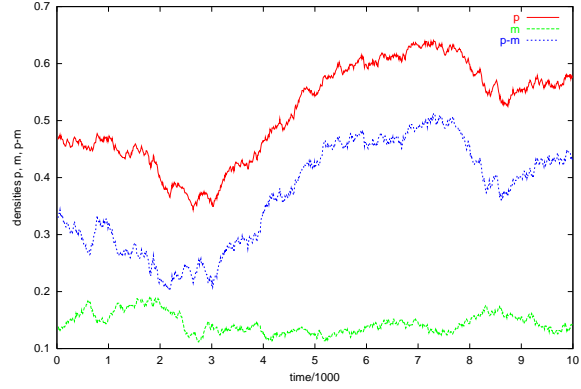}
\end{center}
\caption{Time evolution of the particle densities
$p$, $m$ and density difference $p-m$
during a single simulation at $\beta = 0.275$: minority species has low density $\rho_{\rm min}=1-\sqrt{1-\beta} \simeq 0.149$, density of majority species fluctuates strongly}
\label{fluct}
\end{figure}
\begin{figure}[htbp]
        \begin{center}
                \includegraphics[height=78mm,angle=-90]{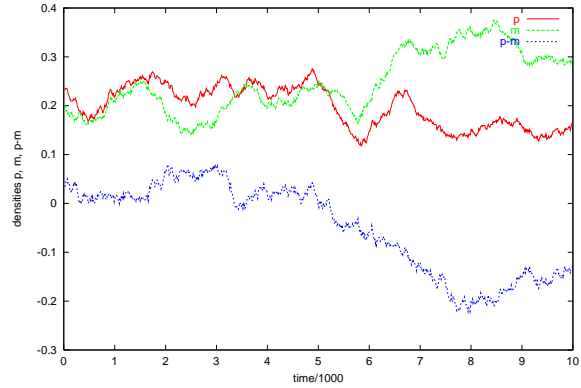}
\end{center}
\caption{
Time evolution of the particle densities
$p$, $m$ and density difference $p-m$
during a single simulation at $\beta = 0.275$: 
  weak fluctuations around $\rho_{\rm sym} =\frac{\beta}{1+\beta}=
  0.216$ turn into asymmetric phase with $\rho_{\rm min} \simeq
  1-\sqrt{1-\beta} = 0.149$ and $\rho_{\rm maj} \simeq \beta = 0.275$}
\label{split}
\end{figure}
\begin{figure}[htbp]
        \begin{center}
          \includegraphics[height=78mm,angle=-90]{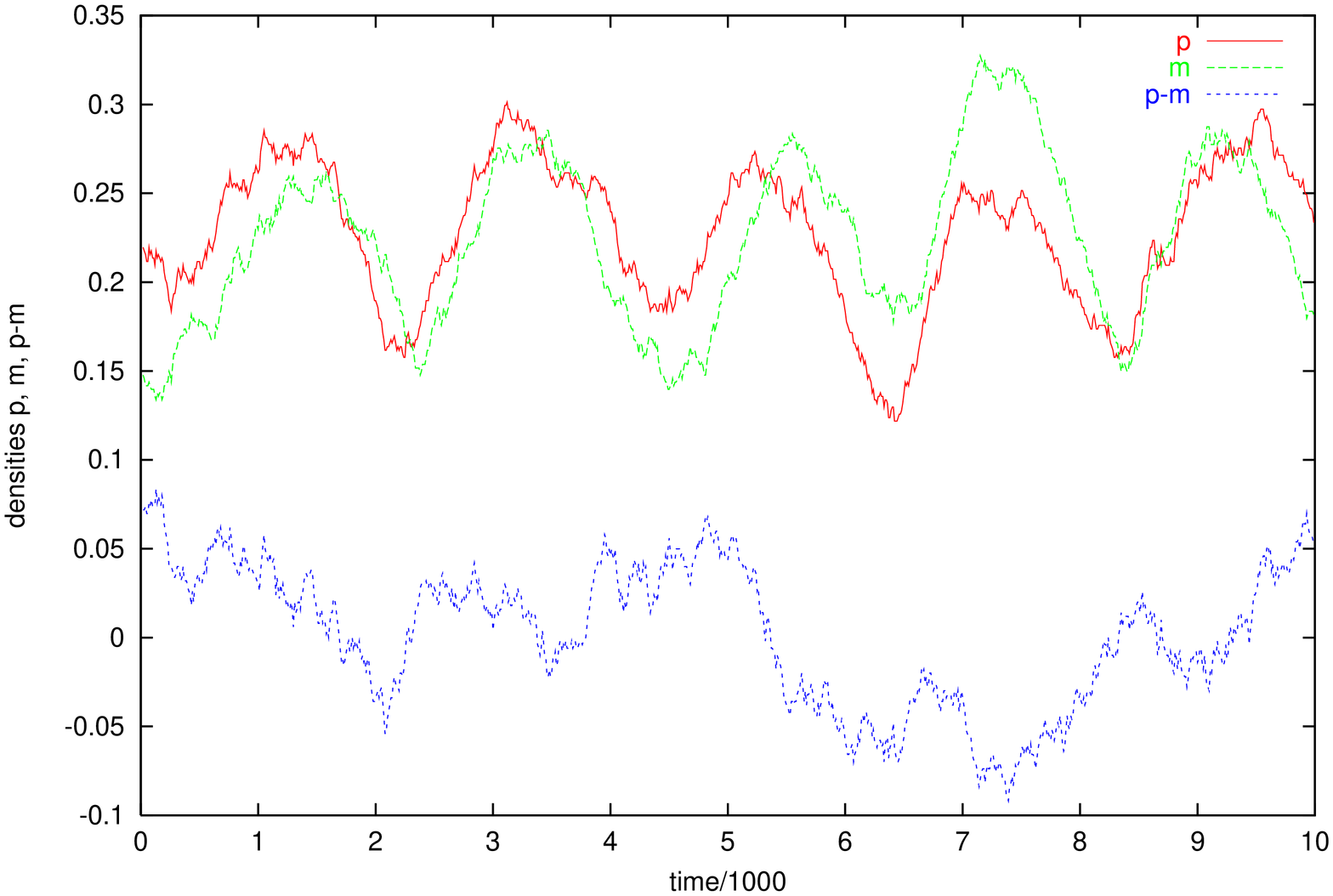}
\end{center}
\caption{
Time evolution of the particle densities
$p$, $m$ and density difference $p-m$
during a single simulation at $\beta = 0.275$: both species fluctuate strongly around density value of symmetric phase $\rho_{\rm sym} =\frac{\beta}{1+\beta} = 0.216$}
\label{oscil}
\end{figure}

We defer the introduction of a more intuitive description than mean-field 
theory, which will explain some of the findings described above, to the next
section.
We first have a closer look at how the behaviour
near the transitions changes with increasing system size.

To provide a more convenient two-dimensional 
representation of $P(p,m)$, 
we will plot in the following the maximum of $P(p,m)$
as seen  along diagonals satifying $p-m= constant$. As the 
probability distribution is symmetric 
around the $p=m$, it is sufficient to restrict the plot
to $p-m>0$.

\begin{figure}[htbp]     
        \begin{center}
                \leavevmode             
                
                \epsfig{file=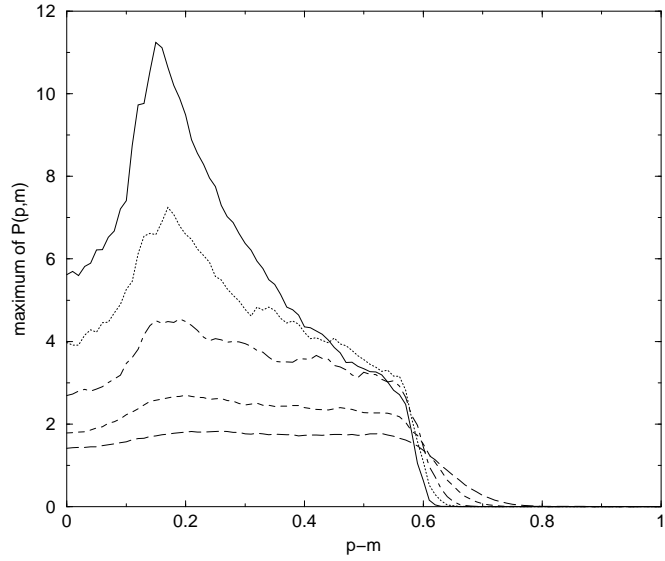,width=75mm,angle=270}
\end{center}
\caption{Maximum of $P(p,m)$ for $\beta=0.275$ and $N=100$, $200$, $500$, $1000$, $2000$}
\label{ld_phase}
\end{figure}

\begin{figure}[htbp]     
        \begin{center}
                \leavevmode             
                
                \epsfig{file=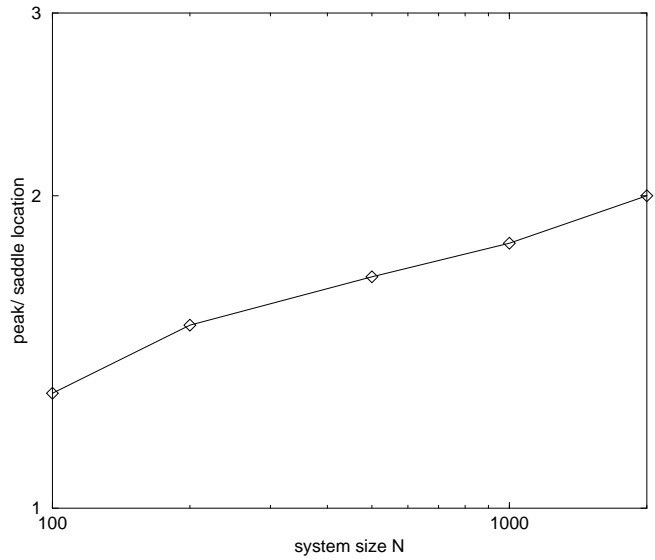,width=75mm,angle=270}
\end{center}
\caption{maximum/saddle ratio in the probability distribution $P(p,m)$}
\label{peak_saddle}
\end{figure}

For increasing system size $N$, the peaks corresponding
to a {\it ld/ld} 
asymmetric region become more pronounced as is shown in Figure \ref{ld_phase} 
for system sizes of $N=100$---$2000$ and for an exit rate 
$\beta = 0.275$. 
To quantify the sharpening of the peaks
we plot the ratio of the height of the peak to the height of saddle
in Figure 
\ref{peak_saddle}. 
Although the increase is slow, it is monotonic and
can be fitted by  a power law 
($\propto N^{0.12}$). This indicates that 
the peaks corresponding to
a {\it ld/ld} phase should dominate
the probability distribution in the large $N$ limit.
Figure~\ref{peak_saddle} thus provides strong evidence 
that the {\it ld/ld} phase does truly exist.

However, one should note that the power-law scaling of the 
peak to saddle
ratio is unusual---within an equilibrium free energy picture
(where $P(p,m) \sim \exp N f(p,m)$)
one would expect the ratio to scale exponentially with $N$.
Conversely the power-law scaling implies that the escape time
or flip time from the {\it ld/ld} phase does not increase
exponentially with $N$. We will return to this point in the conclusion.

\section{Blockage picture}
In the previous section the simulations provided evidence for a {\it
  ld/ld} asymmetric phase but with values of the densities different
from those predicted by mean-field theory.  We also saw coexistence
between the {\it hd/ld} and {\it ld/ld} phases at the transition.

Here we draw on the simulations to give a more intuitive description
of the {\it ld/ld} phase that we will refer to as the blockage
picture. This picture builds on observations of \cite{EFGM} and
\cite{AHR},  the exact results for $\beta\to 0$ \cite{GLEMSS}
and general theoretical considerations of \cite{KSKS,HKPS}.

\begin{figure}[htb]     
        \begin{center}
                \leavevmode             
                \epsfig{file=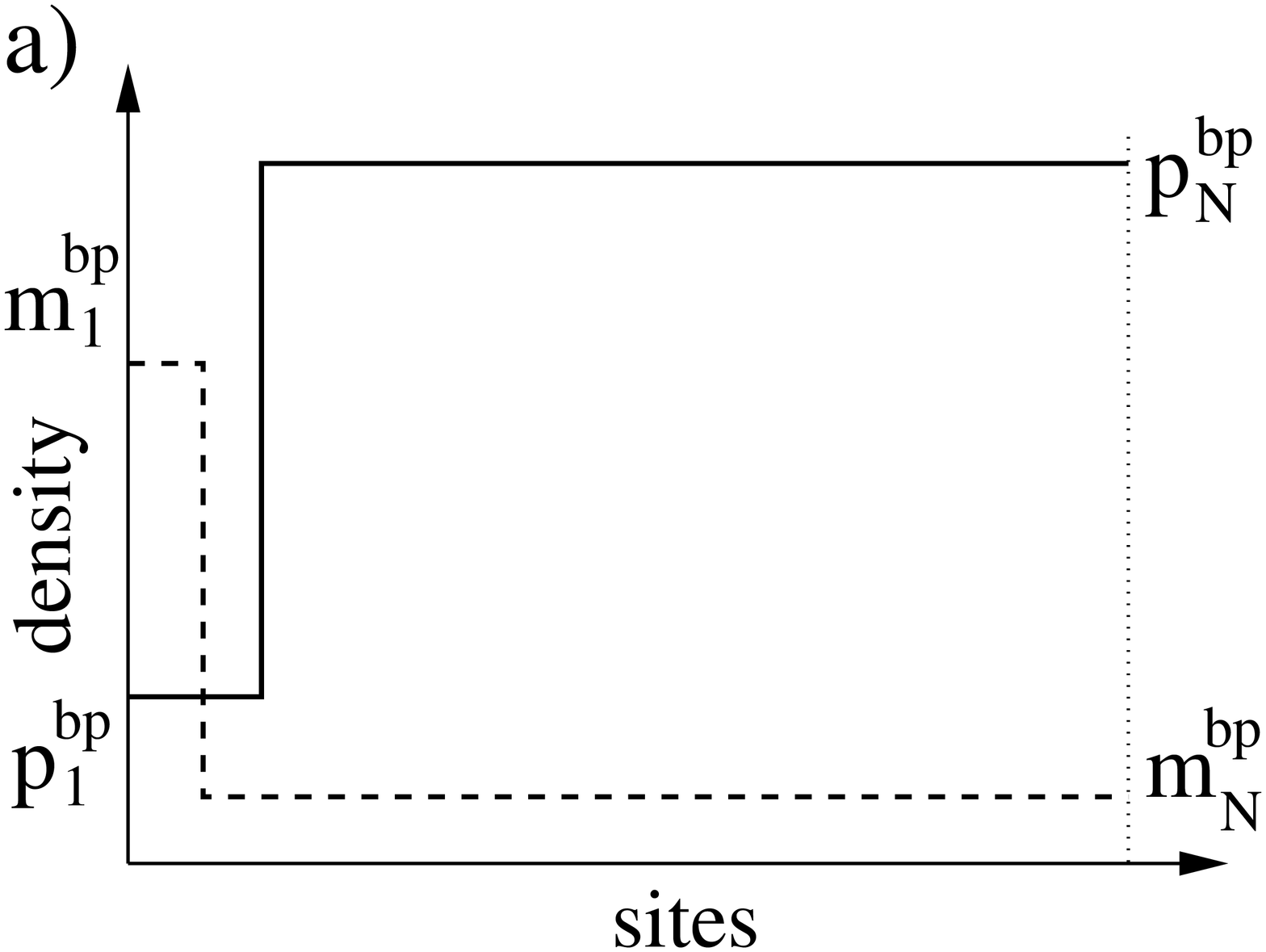,width=50mm,angle=270}
                \hspace*{1cm}
                \epsfig{file=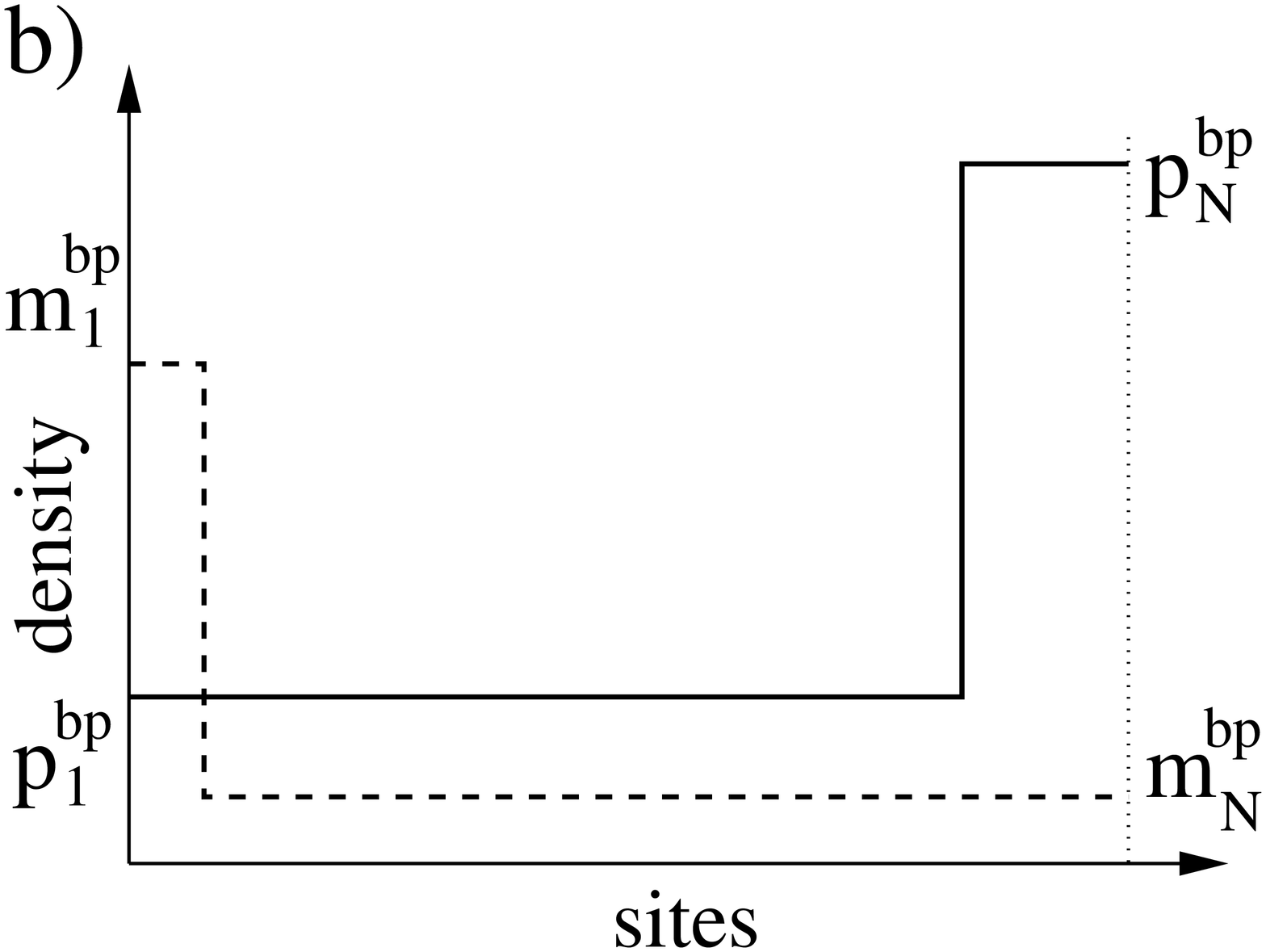,width=50mm,angle=270}
\end{center}
\caption{Schematic picture of the instantaneous
 density profiles in asymmetric phases with $p$ being
 the majority and $m$ being the minority species. a)High density/low
 density b) low density/low density}
\label{asym_dens}
\end{figure}

First we sketch the qualitative features of the {\it hd/ld} phase as
predicted by mean-field theory and confirmed by the exact results for
$\beta\to 0$. Schematically the instantaneous picture is as in
Fig.~\ref{asym_dens}~a): the bulk of the lattice is occupied by a
blockage of positive particles.  In the figure $p_N^{\rm bp}$ denotes
the density of the majority species (travelling from left to right) in
the blockage and $p_1^{\rm bp}$ denotes the density to the left of the
blockage (the upper index `bp' denotes blockage picture).  Similarly
$m_1^{\rm bp}$ denotes the density of the minority species near the
left boundary and $m_N^{\rm bp}$ denotes the density away from the
left boundary. Near the left boundary there is a
small blockage of negative particles. This blockage is unstable in the
sense that the domain wall between it and the bulk region drifts to
the left. Averaging over the positions of the domain wall results in
the exponential decay of the mean field density profile from the left
boundary \cite{KSKS}.

Now we summarise some important features of the {\it ld/ld} phase that
we have observed in simulations.  The simulations show that small
blockages are formed by the majority species at its output end. This
prevents the other particle species from entering and leads to the
symmetry breaking.  However for finite size systems these blockages
are not stable for long times (at least not for times exponential in
the system size) which contrasts with the {\it hd/ld} phase.  

These observations lead us to the following blockage picture for the
{\it ld/ld} phase as illustrated in Figure~\ref{asym_dens}~b).  There
is a small blockage of positive particles at the right end of the
system. Thus the domain wall of Figure~\ref{asym_dens}~a) is now
pushed to the right end of the system.  In order to obtain the
currents and the densities in the {\it ld/ld} phase we assume that the
blockage of the majority species is stable in the sense that the
current into the blockage equals the current out.  We also assume that
the the densities of both species are basically constant within each
domain. This is backed up by our numerical results.  This assumption
of a stable blockage and constant density at the output side of the
majority species implies using the mean-field equations
(\ref{jp_bulk}--\ref{jm_boundary}) that
\begin{equation}
j^{+}= p_N^{\rm bp}(1- p_N^{\rm bp}) = \beta p_N^{\rm bp}
\end{equation}
thus
\begin{equation}
p_N^{\rm bp} = 1-\beta \quad \mbox{and} \quad j^+=\beta(1-\beta)\;.
\label{jpbp}
\end{equation}
Since the current into the blockage is equal to
the current out  one obtains
\begin{equation}
p_1^{\rm bp} = \beta\;.
\label{llrhop}
\end{equation}
The
blockage also controls the input of the minority species and we can 
determine $m_N^{\rm bp}$ from the condition
\begin{equation}
j^-= 1-m_N^{\rm bp}-p_N^{\rm bp} = m_N^{\rm bp}(1-m_N^{\rm bp})
\end{equation}
and one finds 
\begin{equation}
m_N^{\rm bp} = 1-\sqrt{1-\beta}\;.
\label{llrhom}
\end{equation}
Note that these results for the bulk densities
(\ref{llrhop},\ref{llrhom})
differ from the mean-field theory predictions
(\ref{pmft},\ref{mmft}). Also observe
that (\ref{jpbp},\ref{llrhom})
are the same as the corresponding mean-field expressions
for the {\it hd/ld} phase (\ref{mfdensity}).
Thus one can think of the blockage picture as being an
extension of the mean-field theory into the {\it ld/ld} regime.
Finally observe that the majority and minority densities given
by the blockage picture for the {\it ld/ld} phase do not coincide
(except at $\beta=1$ or 0) thus the blockage picture is consistent with a
discontinuous transition to the symmetric phase at some lower value of
$\beta$.

To summarise, in the case of
the {\it hd/ld} phase the blockage occupies the bulk of
the system, the bulk density for the majority species is given by
$\rho^{+} = p_N^{\rm bp} = 1- \beta$. The bulk density for the
minority species is given by $\rho^{-}= m_N^{\rm bp} = 1- \sqrt{1- \beta}$.
At the transition from the {\it hd/ld} to the {\it ld/ld} phase
the domain wall may be found anywhere in the system.  This corresponds
to a shock, that is a sudden change in density over a microscopic
region. The wandering of the shock produces the arms of the boomerang
on the plots of $P(p,m)$ presented in Figure~\ref{transition}.
In the {\it ld/ld} phase the domain wall is localised at the right end
of the system. The bulk densities are given by
(\ref{llrhop},\ref{llrhom})
as
$\rho^{+} = \beta$ and
$\rho^{-}= 1- \sqrt{1- \beta}$.

The blockage picture explains the qualitative features of the phases
and predicts quantitatively the densities in the {\it ld/ld} phase.
However, it does not make any further quantitative predictions
e.g. for the values of $\beta$ at the transition points.  Also some
features remain unexplained.  For example in the {\it ld/ld} phase a
stable blockage (as defined above) is assumed at the right hand
boundary. For such a blockage, one would expect a diffusive motion of
the domain wall leading it to explore the whole system. Yet, this only
occurs at the transition from {\it hd/ld} to {\it ld/ld}; in the {\it
ld/ld} phase the blockage is pinned near that boundary. This suggests
that the blockage is only stable over short times. It would be
interesting to understand this more fully.

\section{Conclusions}
In this paper we studied a totally asymmetric simple exclusion
process for two species which exhibits spontaneous symmetry breaking.
In particular we have made a detailed study of the transition from the
symmetry broken regime to the symmetric regime.  We carried out Monte
Carlo simulations to investigate the transition between the
previously studied high-density/low density asymmetric phase
\cite{EFGM,GLEMSS} and a symmetric phase.  We found evidence for the
existence of a second asymmetric phase.  However although the observed
particle densities in this region are unequal and both smaller than
$1/2$, they do not correspond to the predictions of the mean-field
{\it ld/ld} asymmetric phase.

Instead a simplistic description of this phase is provided by the `blockage 
picture'. It ascribes the symmetry breaking in the two species system 
to the buildup of blockages at the output end of one particle type due 
to the low output rate. These blockages then prevent the other particle 
species from entering which results in a lower density. This picture is 
in accord with mean-field theory for the {\it hd/ld} phase; it 
gives new insight into the observed {\it ld/ld} asymmetric phase.

Interestingly, although both transitions (from {\it hd/ld}
to {\it ld/ld} and from {\it ld/ld} to symmetric) are discontinuous,
they are of different types.
At the first transition one has coexistence between the
{\it hd/ld}
and {\it ld/ld} phases. This is manifested by the presence of a shock
in the density of the majority species wandering through the system.
At the {\it ld/ld} to symmetric transition, however, one cannot have
coexistence simply because a phase with symmetric currents of particles
cannot coexist in the same system
with a phase where the currents are not symmetric.
Thus we have a discontinous transition without coexistence.

We now address the discrepancies between our conclusions
and those of Arndt {\it et al}.

The approach of Arndt {\it et al}\cite{AHR} is to assign a free energy
density to the probability distribution of the steady states.  It is
defined as the negative logarithm of a probability distribution
$P(p-m)$ which depends only on the density difference (and implicitly
on the system size $N$):
\begin{eqnarray}
f(p-m) = \lim_{N\to\infty}\left(-\frac{1}{N}\log{P(p-m)}\right)\label{fe}
\end{eqnarray}
However in that work $P(p-m)$ is calculated by integrating $P(p,m)$
along diagonals satisfying $p-m=constant$. $P(p-m)$ therefore does not
distinguish between the rather narrow peaks and the lower but wider
saddle between them.  Considering the logarithm of this function to
calculate the free energy will flatten any remaining maxima of
$P(p-m)$ in the {\it ld/ld} phase even further. This explains why
Arndt {\it et al} concluded the {\it ld/ld} phase was absent. In
Fig.~\ref{ld_phase}, on the other hand, we project $P(p,m)$ onto a two
dimensional representation by taking the {\rm maximum} along the
diagonals $p-m=constant$.  For the finite size systems considered this
preserves more faithfully the three dimensional structure of $P(p,m)$
and we clearly see the {\it ld/ld} phase.

Finally, we would like to mention the flip time in the asymmetric
phases.  For finite system sizes the blockages in the {\it ld/ld}
asymmetric phase are only stable for short times implying that the
flip time in this asymmetric phase is in turn small.  For $\beta$
vanishingly small (i.e. in the {\it hd/ld} phase) it is known that the
flipping time grows exponentially with $N$ \cite{GLEMSS}, whereas in
the symmetric phase the flipping time increases linearly with $N$.
Both of these dependences have been confirmed by simulations which we
do not present here.  It would be of interest to determine the scaling
of the flip time with $N$ in the {\it ld/ld} phase.  A dependence on
$N$ distinct from the {\it hd/ld} and symmetric phase would provide
further understanding of the {\it ld/ld} phase.  In the {\it ld/ld}
regime, however, the instabilities of the blockages make it difficult
to observe distinct non-linear behaviour for system sizes accessible
by simulations.  We found instead that the {\it ld/ld} phase is
strongly dominated by symmetric behaviour for small systems.  This
simply reflects that the maxima of the probability distribution are
not very pronounced for the {\it ld/ld} asymmetric phase. A
peak-to-saddle ratio of one order of magnitude which should be a
threshold for a crossover to such non-linear behaviour corresponds to
system sizes several orders of magnitude larger than the ones
investigated here.  It remains a numerical challenge to go to such
large system sizes.

{\bf Acknowledgments:} We thank P. Arndt and V. Rittenberg for useful
discussions. MC would like to thank EPSRC and the Gottlieb Daimler -
und Karl Benz - Stiftung for financial support. The support of the
Israeli Science Foundation is gratefully acknowledged (DM).

\end{document}